\documentclass[12pt,a4paper]{article}
\usepackage{graphicx}
\usepackage{longtable}
\usepackage{wrapfig}
\usepackage{rotating}
\usepackage[normalem]{ulem}
\usepackage{capt-of}
\usepackage{hyperref}
\usepackage{amsmath}
\usepackage{amssymb}
\usepackage{mathtools}
\usepackage{mathrsfs}
\usepackage{float}
\usepackage{subcaption}
\usepackage[version=4]{mhchem}
\usepackage[x11names]{xcolor}
\usepackage{indentfirst}
\usepackage{geometry}
\usepackage{placeins}
\usepackage{setspace}
\usepackage[bracket-numbers = false, table-align-uncertainty = false, table-align-exponent=true]{siunitx}
\usepackage{xurl}
\usepackage{amsmath}
\usepackage{pdfpages}
\usepackage{anyfontsize}
\usepackage{pdflscape}
\usepackage[english]{babel}
\usepackage{tikz}
\usetikzlibrary{shapes,arrows,shapes.multipart,mindmap,trees,overlay-beamer-styles}
\usepackage[absolute,overlay]{textpos}
\TPGrid[0mm,0mm]{100}{100}
\usepackage{authblk}
\author[1]{Diogo P.~L.~Carvalho}
\author[1,2]{Ana C.~B.~Loponi}
\author[1,3]{Daniel R.~Cassar}
\affil[1]{Ilum School of Science, Brazilian Center for Research in Energy and Materials (CNPEM), Zip Code 13083-970, Campinas, Sao Paulo, Brazil.}
\affil[2]{Brazilian Biorenewables National Laboratory, Brazilian Center for Research in Energy and Materials (CNPEM), Zip Code 13083-970, Campinas, Sao Paulo, Brazil.}
\affil[3]{National Institute of Science and Technology on Materials Informatics, Campinas, Sao Paulo, Brazil}
\affil[ ]{\small\texttt{daniel.cassar@ilum.cnpem.br}}
\date{}
\title{Will it form a glass? Tackling glass formation using binary classification}
\hypersetup{
 pdfauthor={Daniel R. Cassar},
 pdftitle={Will it form a glass? Tackling glass formation using binary classification},
 pdfkeywords={},
 pdfsubject={},
 pdfcreator={Daniel Roberto Cassar}, 
 pdflang={English},
 pdfborder={0 0 0},
 pdfborderstyle={},
 colorlinks=true, urlcolor = DodgerBlue4, citecolor = DodgerBlue4, linkcolor = DodgerBlue4}
\usepackage{calc}
\newlength{\cslhangindent}
\newlength{\csllabelsep}
\newlength{\csllabelwidth}
\newenvironment{cslbibliography}[2] % 1st arg. is hanging-indent, 2nd entry spacing.
 {% By default, paragraphs are not indented.
  \setlength{\parindent}{0pt}
  \ifodd #1
  \let\oldpar\par
  \def\par{\hangindent=\cslhangindent\oldpar}
  \fi
  \setlength{\parskip}{\parskip +  #2\baselineskip}
 }%
 {}

\newcommand{\cslleftmargin}[1]{\parbox[t]{\csllabelsep + \csllabelwidth}{#1}}
\newcommand{\cslrightinline}[1]
  {\parbox[t]{\linewidth - \csllabelsep - \csllabelwidth}{#1}\break}

\newcommand{\cslbibitem}[2]
  {\leavevmode\vadjust pre{\hypertarget{citeproc_bib_item_#1}{}}#2}
\makeatletter
\newcommand{\cslcitation}[2]
 {\protect\hyper@linkstart{cite}{citeproc_bib_item_#1}#2\hyper@linkend}
\makeatother
\begin{document}

\maketitle
\vspace{-0.5cm}

\noindent\fbox{\begin{minipage}[t]{1\columnwidth - 2\fboxsep - 2\fboxrule}%
\begin{center}This manuscript has been submitted for peer review and has not yet been formally published. As citation details (e.g., journal, DOI, volume) may be updated upon acceptance, readers are encouraged to consult the arXiv preprint page for the current recommended citation. \end{center}
\end{minipage}}

\vspace{0.5cm}
\section*{Abstract}
\label{sec:org3fb9547}
Glass formation is one of the most important and fundamental open problems in glass science. Predicting whether a liquid can be easily frozen into a glass appears simple but is far from it. In this communication, we address glass formation in inorganic nonmetallic liquids using binary classification to predict the probability that a given liquid will form a glass under typical laboratory conditions. Using a dataset of more than 50,000 examples, we trained random forest classifiers that achieved ROC-AUC values around 0.89 and PR-AUC close to 0.95 on the holdout dataset (i.e., unseen data). A rigorous model selection routine was employed, including hyperparameter tuning with cross-validation, and four different data treatment routes were evaluated. Using SHAP values, we extracted valuable insights from the trained models that both agree with established knowledge and extend it. For example, we identified that the bandgap energy of the constituent chemical elements is positively correlated with glass formation. When glass stability parameters and Jezica were added to the dataset, no performance improvement was observed, but model complexity decreased significantly. This result is particularly relevant for composition screening, especially in inverse design problems.

\newpage
\section{Introduction}
\label{sec:org7a1bd38}

Glasses are noncrystalline materials that exist out of thermodynamic equilibrium \cslcitation{1}{[1]}. The vast majority of glasses are produced via melt-quenching processes, wherein a liquid is cooled rapidly enough to bypass crystallization below the liquidus temperature. Glass formation requires a delicate balance between kinetic and thermodynamic factors, a fundamental complexity captured in Philip Anderson's observation that ``the deepest and most interesting unsolved problem in solid state theory is probably the theory of the nature of glass and the glass transition'' \cslcitation{2}{[2]}.

The ease of glass formation varies dramatically across different liquid systems. Outstanding glass formers such as \ce{B2O3}, anorthite, and atactic polystyrene readily vitrify upon cooling and are so resistant to crystallization that inducing crystalline phases in these materials poses significant experimental challenges \cslcitation{3}{[3]}. Conversely, extremelly resilient glass formers like water and the vast majority of metallic liquids require extraordinarily high cooling rates to preserve their disordered liquid structures in the glassy state \cslcitation{4}{[4]}. Consequently, glass forming ability (GFA) is commonly quantified as the inverse of the critical cooling rate (that is, the minimum cooling rate required to avoid crystallization during vitrification) \cslcitation{5}{[5]}.

Predicting GFA remains a formidable challenge given the unresolved nature of the glass transition itself. Numerous studies have addressed this problem for both nonmetallic \cslcitation{5}{[5]}, \cslcitation{6}{[6]}, \cslcitation{7}{[7]}, \cslcitation{8}{[8]}, \cslcitation{9}{[9]}, \cslcitation{10}{[10]}, \cslcitation{11}{[11]} and metallic systems \cslcitation{12}{[12]}, \cslcitation{13}{[13]}, \cslcitation{14}{[14]}, \cslcitation{15}{[15]}, \cslcitation{16}{[16]}, \cslcitation{17}{[17]}, \cslcitation{18}{[18]}, \cslcitation{19}{[19]}, \cslcitation{20}{[20]}, \cslcitation{21}{[21]}. In this work, we simplify the prediction task to a binary classification problem: given an inorganic liquid composition, will it form a glass under standard laboratory cooling conditions?
\section{Materials and methods}
\label{sec:org4ce80ff}
\subsection{Data aquisition and processing}
\label{sec:org215c34a}

Experimental data for this study were extracted from the SciGlass database \cslcitation{22}{[22]}. The dataset comprises unary, binary, and ternary inorganic glass compositions, with concentrations reported in mol percentage. Each composition is labeled with a text string describing the synthesis outcome reported in the original literature source. Representative labels include ``Glass'', ``Crystal'', ``Clear glass'', and numerous variations thereof. 

The label-processing step involved filtering instances whose labels unambiguously correspond to glass formation or crystallization within the binary classification framework described in the introduction. \uline{Positive} class assignment was applied to instances with the following labels: Glass, Clear glass, Phase-separated glass, Stable glass, Phase separated glass, Homogeneous glass, Transparent glass, Glassy phase, glass, Pure glass, Stable glasses, Clear glass (cast), and Glasses. \uline{Negative} class assignment was applied to instances labeled as: Crystal, Crystalline, Crystallization, Fully devitrified glass, crystal, Crystallized, Crystall, and Grystal. All instances bearing alternative labels were excluded from further analysis.

After that, the dataset was converted from compound mole percentage to atomic fraction. A unimodal deduplication routine was applied to address redundant entries. This routine was based exclusively on the elemental fraction features, considering three decimal places. When duplicate compositions exhibited bimodal class distributions (equal numbers of positive and negative labels), all associated instances were removed due to label ambiguity. Conversely, when duplicates exhibited unimodal class distribution, they were collapsed into a single entry assigned the modal class label.

The resulting dataset, hereafter referred to as the CHEM dataset, contains atomic fraction compositions as features and binary glass formation outcomes as targets.
\subsection{Feature engineering}
\label{sec:org3902471}

Feature engineering was performed on the CHEM dataset using the \texttt{GlassPy} library \cslcitation{23}{[23]}. This process generated a set of physicochemical descriptors based on atomic fraction information. The underlying assumption is that enriching the feature space with domain-specific information enables machine learning algorithms to induce more accurate predictive models.

The feature engineering protocol followed the methodology detailed in Ref. \cslcitation{24}{[24]}. Briefly, this method computes the Hadamard product (element-wise multiplication) between the atomic fraction vector and a physicochemical property vector (e.g., atomic radius, electronegativity, melting point, and others). The resulting vector is reduced to a scalar value via an aggregation function; specifically, the sum, mean, minimum, maximum, and standard deviation were employed in this work. This process yields what are termed \emph{weighted descriptors}. Alternatively, applying a ceiling operator to the molar fraction vector prior to the Hadamard product generates \emph{absolute descriptors}. The rationale is that weighted descriptors assist the model in the classification task, whereas absolute descriptors help identify distinct glass families through clustering. Readers are directed to Ref. \cslcitation{24}{[24]} for comprehensive technical details.

In addition to physicochemical descriptors, two metadata features were incorporated: the total number of distinct chemical elements in the composition and the total count of components in the formulation.

This process produced approximately 600 distinct features, encompassing atomic fractions, physicochemical descriptors, and metadata, thereby presenting a high-dimensional modeling challenge. To address this, Sequential Forward Selection (SFS) was employed for feature selection \cslcitation{25}{[25]}.

SFS is an iterative algorithm that begins with an empty feature set and sequentially adds features one at a time. In each iteration, the algorithm selects the feature that yields the highest performance improvement when combined with the existing feature set. Performance was evaluated using the minimum of the mean and median macro-averaged F1-scores, computed across all folds of a stratified 10-fold cross-validation. This metric provides a conservative and robust estimate, ensuring balanced predictive capabilities across both classes.

The selection process continued until a stopping criterion of 100 features was reached. Although SFS identified an ordered subset of 100 relevant descriptors, only the top 20 were retained for subsequent analysis, as no significant performance gains were observed beyond this point. The resulting dataset, containing the selected engineered features and binary glass formation outcomes, is hereafter referred to as the FEATENG dataset.
\subsection{Estimating glass stability parameters}
\label{sec:org11fdafb}

An additional class of descriptors relevant to GFA prediction comprises glass stability parameters \cslcitation{6}{[6]}. These parameters are typically computed from characteristic temperatures including the glass transition temperature (\(T_g\)), liquidus temperature (\(T_\mathrm{liq}\)), crystallization peak temperature (\(T_c\)), and crystallization onset temperature (\(T_x\)). In this work, we calculated the Weinberg (\(K_{w,T_x}\) and \(K_{w,T_c}\)) \cslcitation{26}{[26]}, Hrübý (\(K_h\)) \cslcitation{27}{[27]}, Saad and Poulain (\(H'\)) \cslcitation{28}{[28]}, and Lu and Liu (\(\gamma\)) \cslcitation{29}{[29]} glass stability parameters, as defined by the equations below. These specific parameters were selected based on their reported strong correlations with experimental GFA measurements \cslcitation{8}{[8]}.

\begin{equation}
K_{w,T_x} = \frac{T_x - T_g}{T_\mathrm{liq}}
\end{equation}

\begin{equation}
K_{w,T_c} = \frac{T_c - T_g}{T_\mathrm{liq}}
\end{equation}

\begin{equation}
K_{h} = \frac{T_c - T_g}{T_\mathrm{liq} - T_c}
\end{equation}

\begin{equation}
H' = \frac{T_x - T_g}{T_g}
\end{equation}

\begin{equation}
\gamma = \frac{T_c}{T_g + T_\mathrm{liq}}
\end{equation}

Additionally, the \emph{Jezica} parameter was computed (see Eq. \ref{eq:jezica}). While not strictly a glass stability parameter, \emph{Jezica} serves as a proxy for GFA \cslcitation{5}{[5]}. In the expression shown in Eq. (\ref{eq:jezica}), \(\eta(T)\) is the equilibrium shear viscosity measured at temperature \(T\).

\begin{equation}
\label{eq:jezica}
\textit{Jezica} = \frac{\eta(T_\mathrm{liq})}{T_\mathrm{liq}^2}
\end{equation}

GlassNet \cslcitation{24}{[24]} was employed to estimate the values of \(T_g\), \(T_\mathrm{liq}\), \(T_c\), \(T_x\), and \(\eta(T_\mathrm{liq})\) for computing the parameters described above. It is recognized that error propagation in these calculations may pose challenges when attempting to directly correlate computed values with experimental GFA \cslcitation{10}{[10]}. Nevertheless, the expectation is that machine learning algorithms can, in principle, leverage this information to enhance predictive performance.

The resulting dataset comprising five glass stability parameters plus \emph{Jezica} as features, together with binary glass formation outcomes as targets, is hereafter referred to as the GS dataset. A merged dataset combining features from both FEATENG and GS datasets is designated as FEATENG+GS.
\subsection{Model selection}
\label{sec:org6795237}

Each of the four datasets (CHEM, FEATENG, GS, and FEATENG+GS) was partitioned into training and holdout subsets using a 90:10 ratio. To ensure consistency, the exact same examples were assigned to the holdout set across all dataset versions.

Following this split, independent model selection was conducted for each dataset. This process involved evaluating 2,000 hyperparameter configurations, sampled from the search space defined in Table \ref{tab:hp_search_space}. The optimization was guided by the Tree-structured Parzen Estimator (TPE) \cslcitation{30}{[30]}, a Bayesian optimization algorithm implemented in the \texttt{optuna} Python module \cslcitation{31}{[31]}.

The optimization objective was to maximize model performance using a conservative cross-validation metric: the minimum of the mean and median macro-averaged F1-scores calculated across 10 folds. This criterion was selected to prioritize robust models that demonstrate stable performance across different data partitions.

Ultimately, the configuration with the highest estimated performance was retained for each dataset. These four selected models were then evaluated on the holdout set to determine their final performance.
\subsection{Explainable AI}
\label{sec:orgd42a5c9}

As random forests do not produce inherently interpretable models, we employed explainable AI techniques to investigate the relationships learned by the algorithm. Specifically, we estimated Shapley values \cslcitation{32}{[32]} for each feature using the \texttt{shap} Python module, computed on the train dataset. These estimated values, referred to as SHapley Additive exPlanations (SHAP) values, quantify the contribution of individual features to each model prediction \cslcitation{33}{[33]}, \cslcitation{34}{[34]}, \cslcitation{35}{[35]}.
\subsection{Use of Artificial Intelligence}
\label{sec:org540c63a}

This paper was originally written in English by the authors and subsequently refined using AI-assisted technologies (Perplexity and Claude) to improve readability, as English is not the native language of the authors. All AI-generated suggestions were reviewed and edited by the authors, who take full responsibility for the content of this publication.
\section{Results and discussion}
\label{sec:orgb066e58}
\subsection{Data analysis and selected phisicochemical features}
\label{sec:org7fe0568}

Following data acquisition and processing, the final dataset comprises 36,033 positive instances (70.8\%, glass-forming compositions) and 14,864 negative instances (29.2\%, non-glass-forming compositions), indicating a moderate class imbalance.

\begin{figure*}[h]
\centering
\includegraphics[keepaspectratio,width=0.9\textwidth]{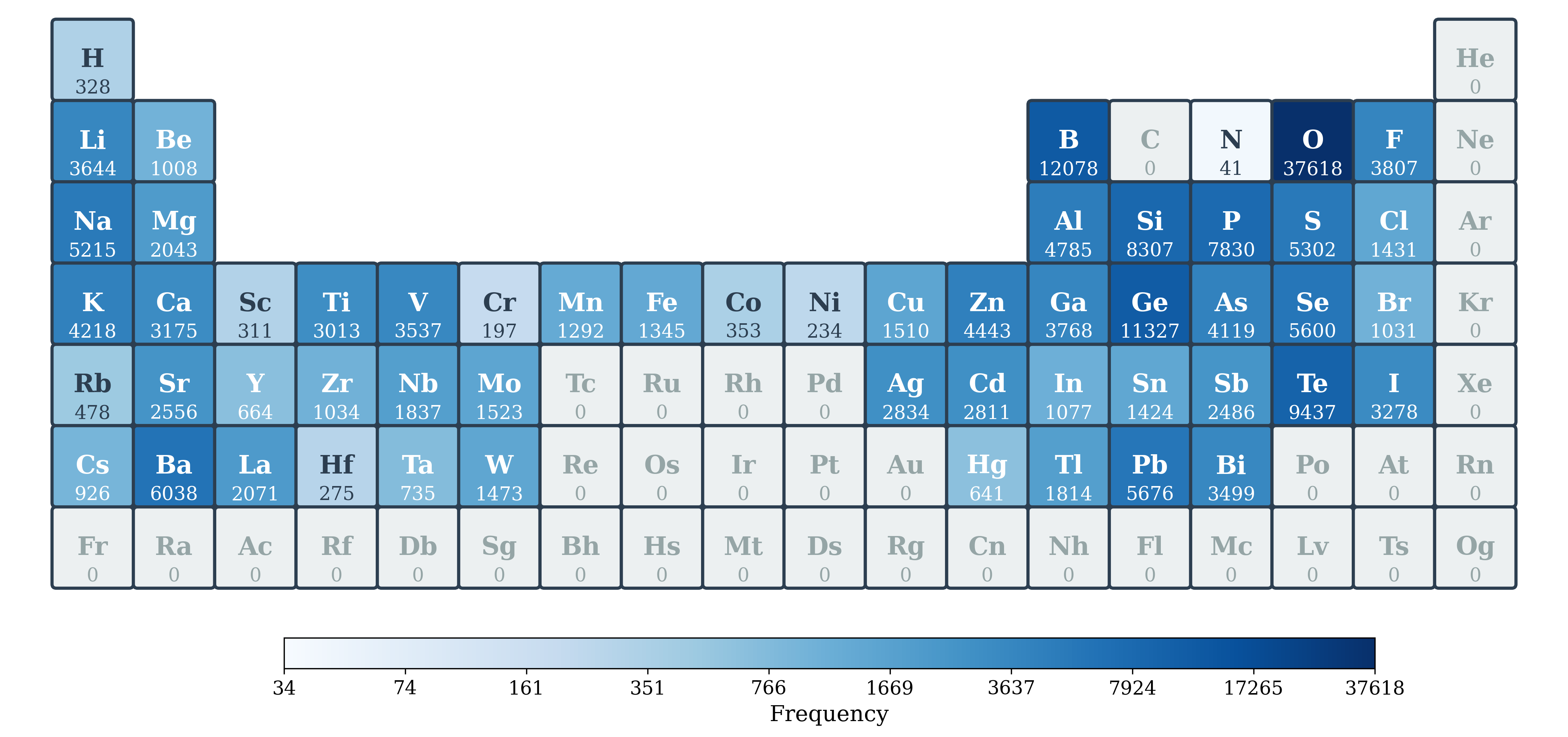}
\caption{\label{fig:periodic_table}Periodic Table heatmap illustrating the frequency of occurrence for each chemical element in the entire dataset, before the holdout split. Color intensity represents frequency on a logarithmic scale.}
\end{figure*}

\begin{figure*}[h]
\centering
\includegraphics[keepaspectratio,width=0.9\textwidth]{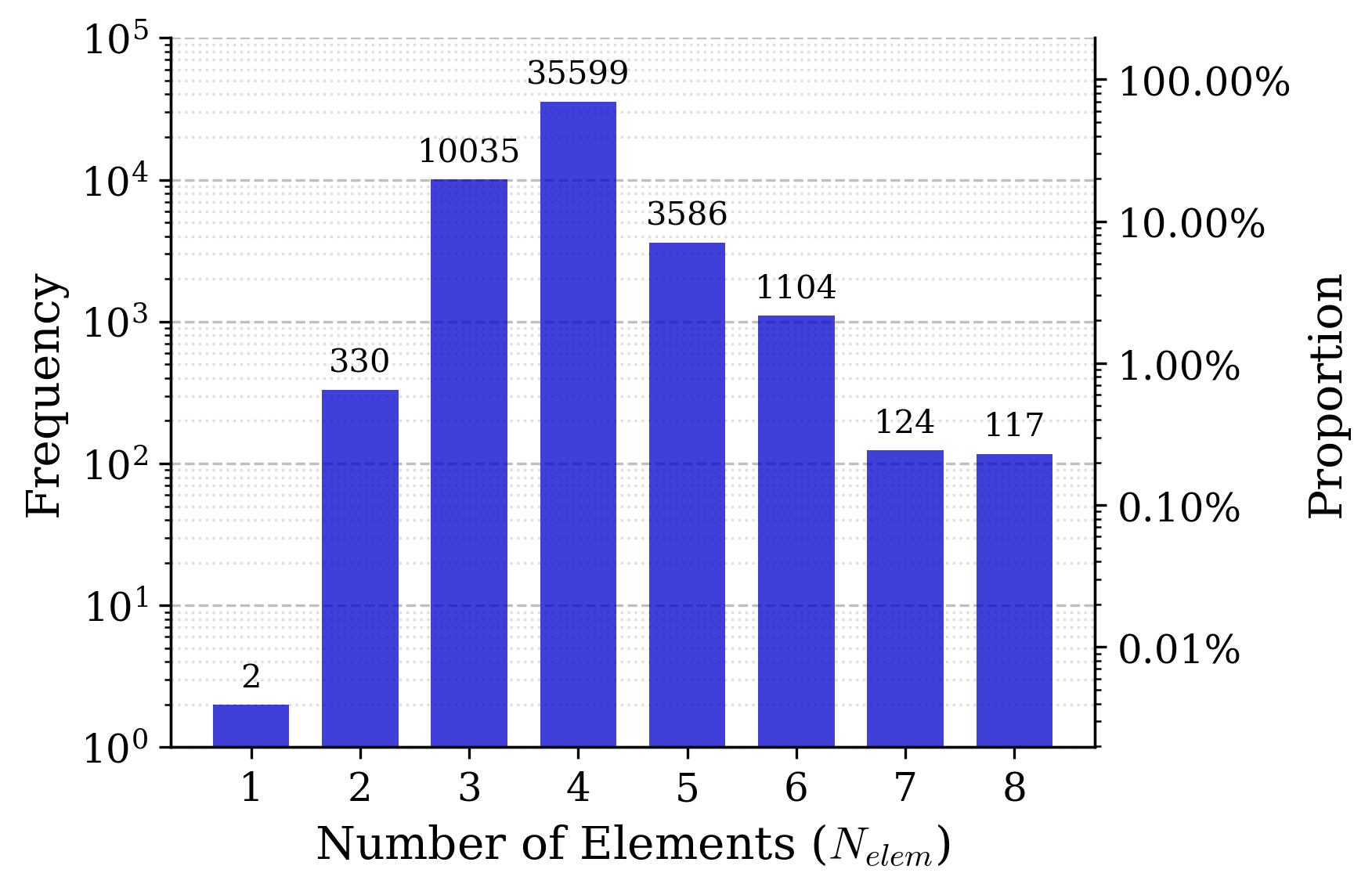}
\caption{\label{fig:num_elements}Distribution of the number of elements in the CHEM dataset.}
\end{figure*}

The CHEM dataset includes 67 distinct chemical features: Ag, Al, As, B, Ba, Be, Bi, Br, Ca, Cd, Ce, Cl, Co, Cr, Cs, Cu, Dy, Er, Eu, F, Fe, Ga, Gd, Ge, H, Hf, Hg, Ho, I, In, K, La, Li, Lu, Mg, Mn, Mo, N, Na, Nb, Nd, Ni, O, P, Pb, Pr, Rb, S, Sb, Sc, Se, Si, Sm, Sn, Sr, Ta, Tb, Te, Ti, Tl, Tm, V, W, Y, Yb, Zn, and Zr. The occurrence frequencies of these elements are highly heterogeneous, as illustrated in Fig. \ref{fig:periodic_table}. Additionally, Fig. \ref{fig:num_elements} shows the distribution of distinct chemical elements per composition ranges from 1 to 8, with the majority of samples concentrated between 3 and 5 elements. Notably, the principle of maximum confusion suggests that increasing the number of elements in a mixture increases the likelihood of glass formation \cslcitation{36}{[36]}. This implies that the dataset covers the most critical compositional region for the problem under investigation. Table \ref{tab:descriptive} complements this analysis by showing the descriptive statistics for all chemical elements, considering the entire dataset.

The FEATENG dataset consists of 20 selected features, comprising 11 weighted descriptors and 9 absolute descriptors. Ranked from highest to lowest importance (acording to the SFS algorithm), these features are: the absolute std. dev. of the Van der Waals radius, weighted mean of the atomic radius, weighted mean of the number of filled p valence orbitals, absolute sum of the Pettifor number, weighted sum of melting enthalpy, weighted mean of the number of oxidation states, absolute mean of melting enthalpy, weighted mean of the number of filled d valence orbitals, weighted sum of the number of filled s valence orbitals, absolute std. dev. of the DFT bandgap energy of the \(T=0\) K ground state, weighted mean of the DFT volume of the \(T=0\) K ground state, weighted sum of the DFT bandgap energy of the \(T=0\) K ground state, absolute std. dev. of the number of oxidation states, weighted maximum of the number of unfilled d valence orbitals, weighted sum of the number of unfilled p valence orbitals, absolute std. dev. of the melting point, absolute sum of the number of unfilled valence orbitals, absolute maximum of the number of oxidation states, weighted std. dev. of the covalent atomic radius, and the absolute std. dev. of electronegativity on the Martynov-Batsanov scale. Table \ref{tab:symbols} shows the symbols and references of these selected features. 

The proposed data-driven approach identified atomic radius, the number of oxidation states, and electronic structure (occupancy of valence orbitals) as critical information for predicting glass formation. Significantly, these findings align with established physical theories linking these properties to Glass Forming Ability (GFA).

The relationship between atomic radius and GFA is well known and has historical roots in Goldschmidt's Radius Ratio Criterion \cslcitation{37}{[37]} and Zachariasen's Random Network Theory \cslcitation{38}{[38]}. Specifically, a cation-to-anion radius ratio between 0.225 and 0.414 promotes low cation coordination, a geometric condition that favors glass formation. Furthermore, atomic radius mismatch is a known determinant in the formation of metallic glasses \cslcitation{4}{[4]}.

The significance of oxidation states correlates with Sun's Bond Strength Model \cslcitation{39}{[39]}. The oxidation state governs the polarization of coordinated oxygen atoms by the cation, thereby influencing both network rigidity and the propensity for crystallization. This interpretation is further supported by Dietzel's field strength concept \cslcitation{40}{[40]}, which observes that small, high-charge cations (associated with high oxidation states) tend to act as glass formers, whereas large, low-charge cations tend to act as modifiers.

The connection between electronic structure (filled vs. unfilled valence orbitals) and GFA is related to Phillip--Thorpe Topological Constraint Theory (TCT) for covalently bonded systems \cslcitation{41}{[41]}, \cslcitation{42}{[42]}, \cslcitation{43}{[43]}. In the framework of TCT, one computes the average coordination number of atoms in the system, \(\left\langle r \right\rangle\), and checks if it is above or below the critical value of 2.4. Above means that the glass network is overconstrained and below means that it is underconstrained, both cases reduces glass forming ability.

Additionally, it is worth mentioning that the melting enthalpy was also identified as a relevant property in our analysis. This quantity is related to the thermodynamic driving force for crystallization \cslcitation{44}{[44]}, \cslcitation{45}{[45]}.
\subsection{Model selection and performance}
\label{sec:orgee972bc}

Following the hyperparameter tuning procedure described in the methodology, Table \ref{tab:hp_search_space} shows the selected hyperparameters for each dataset. The models selected a reasonably high number of estimators, close to 400 for models considering glass stability parameters and close to 700 for others. Since each estimator represents a different decision tree, this high number underscores the difficulty of predicting whether a liquid will form a glass. Interestingly, models with lower complexity (smaller number of estimators) were selected for datasets containing glass stability parameters (datasets GS and FEATENG+GS).

Given the moderate class imbalance in the data, weighted strategies were unsurprisingly selected over uniform weighting. Regarding the split criterion (the function that quantifies split quality for each conditional branch in the decision trees), the Gini impurity was selected for the CHEM dataset, while cross-entropy (logarithmic loss) was selected for other datasets. This difference is hypothesized to stem from the high dimensionality and sparsity of CHEM dataset.

No model imposed a limit on the number of leaf nodes. Limiting leaf nodes is a common regularization strategy that increases bias while reducing variance in decision trees. The absence of this limit suggests that the high complexity of this problem benefits from additional variance rather than increased bias.

Finally, the GS dataset required higher values for both the minimum number of samples to allow splitting and the minimum number of samples per leaf node compared to other datasets. These stricter regularization strategies were likely selected due to the high uncertainty in the data, as previously discussed.
\subsection{Model performance}
\label{sec:org3147cd8}

\begin{figure*}[h]
\centering
\includegraphics[keepaspectratio,width=0.9\textwidth]{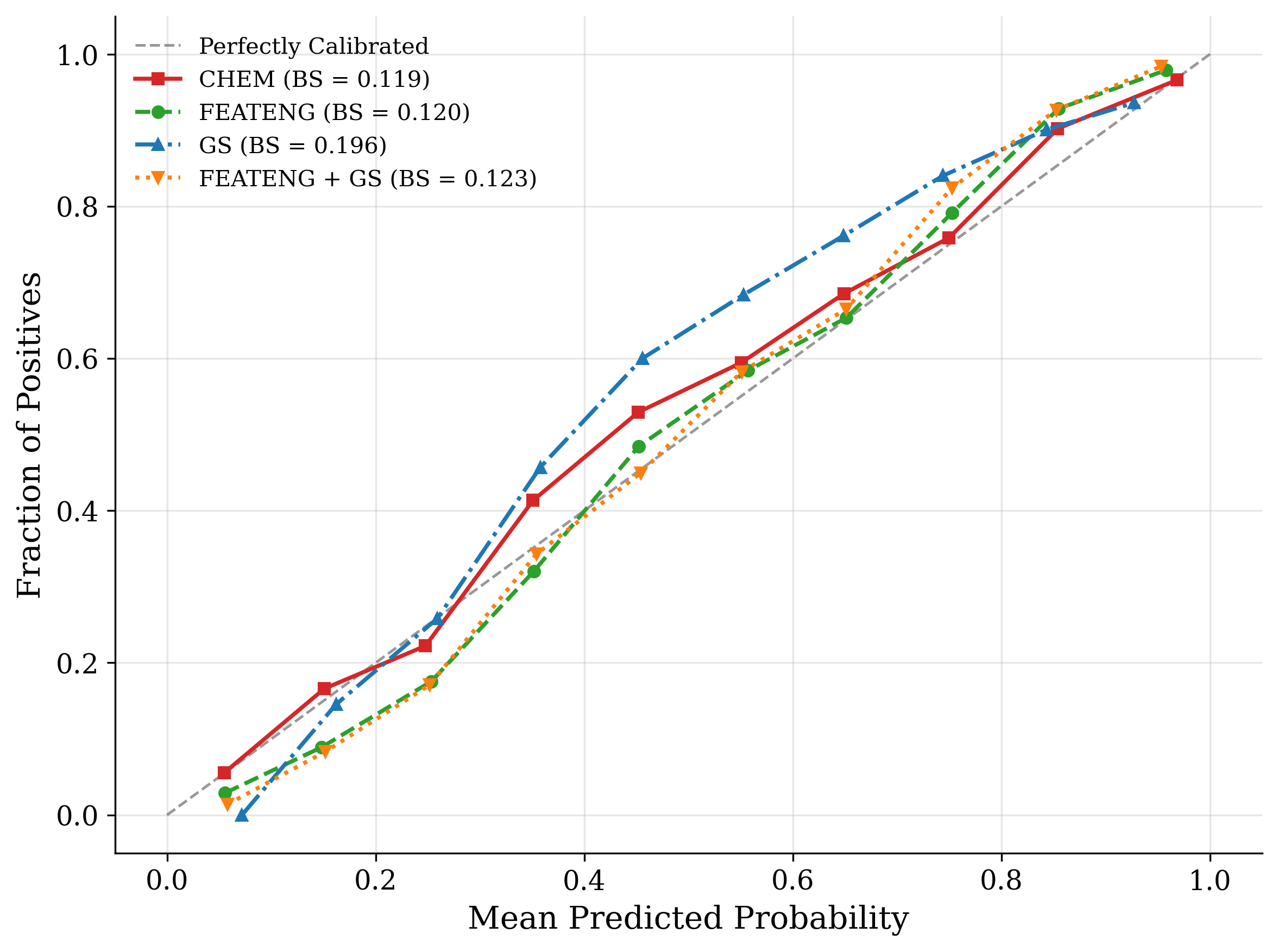}
\caption{\label{fig:cal_curve}Calibration curves for the glass formation models. The plot compares the mean predicted probability against the actual fraction of positive samples in each bin. The diagonal dotted line represents perfect calibration. The Brier Score (BS) is reported for each model, where lower values indicate better probabilistic calibration.}
\end{figure*}

\begin{figure*}[h]
\centering
\includegraphics[keepaspectratio,width=0.9\textwidth]{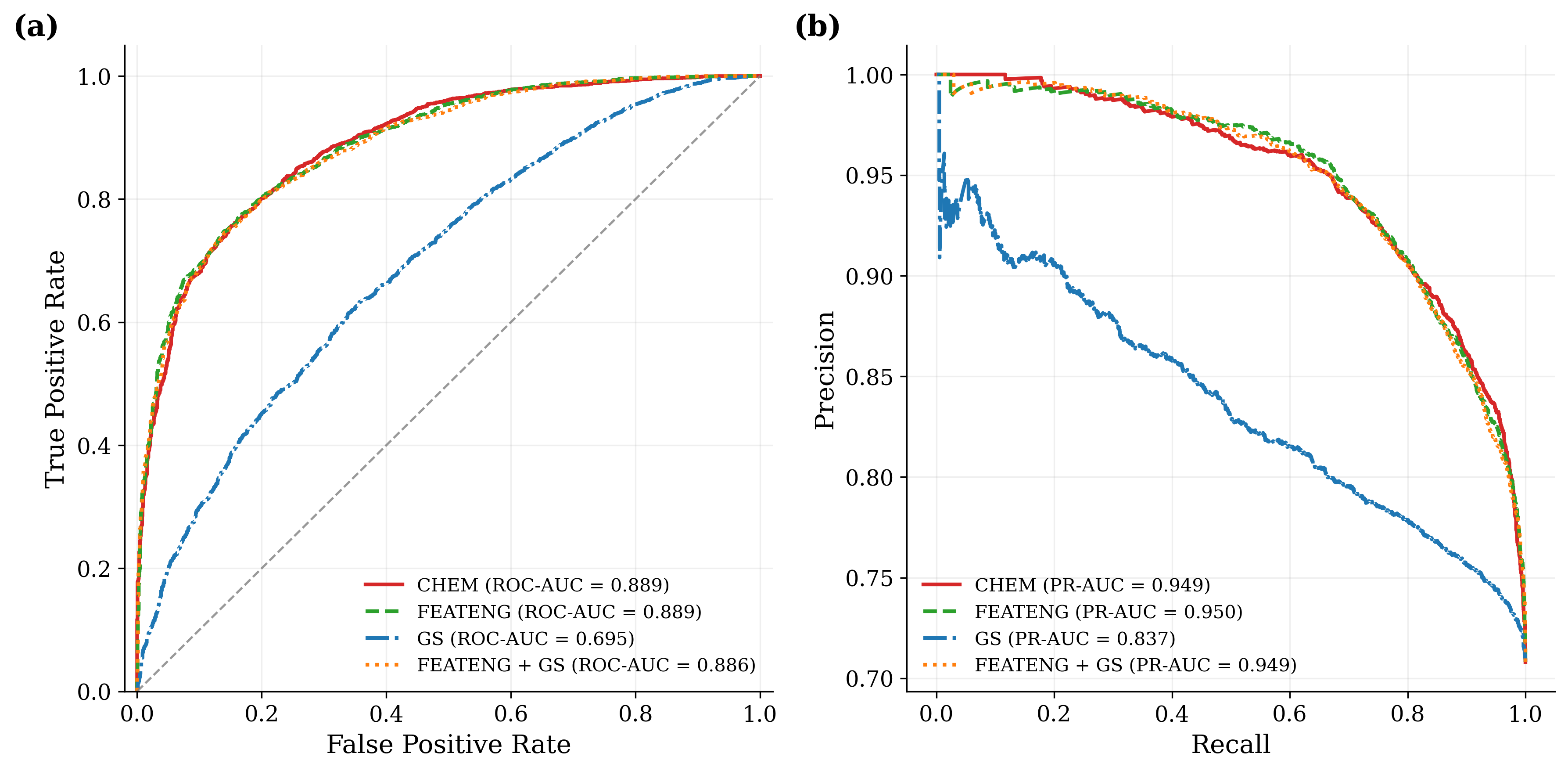}
\caption{\label{fig:roc_pr_auc}Performance comparison of the glass formation models. (a) Receiver Operating Characteristic curves and (b) Precision-Recall curves.}
\end{figure*}

The performance of the models can be studied and analyzed in Figs. \ref{fig:roc_pr_auc}, \ref{fig:cal_curve}, and Table \ref{tab:metrics}. Fig. \ref{fig:roc_pr_auc} compares the area under the curve (AUC) for Receiver Operating Characteristic (ROC) and Precision-Recall (PR) curves across the four models. The models induced from the CHEM, FEATENG, and FEATENG+GS datasets yield nearly indistinguishable performance, with ROC-AUC values around 0.89 and PR-AUC close to 0.95. In contrast, the GS model performs substantially worse in both metrics, achieving a ROC-AUC of about 0.70 and a PR-AUC of about 0.84. This indicates limited discriminative power when relying solely on glass stability parameters. Importantly, these GS parameters were predicted rather than measured, and combining predictions is known to greatly increase error in final calculations \cslcitation{10}{[10]}.

Fig. \ref{fig:cal_curve} shows the calibration curve for all models. Overall, all models exhibit good probabilistic calibration (with predicted probabilities closely following the ideal diagonal), eliminating the need for post-hoc recalibration. The models induced from the CHEM, FEATENG, and FEATENG+GS datasets exhibit very similar behavior across the probability range. However, the model induced from the GS dataset shows slightly larger deviations, consistent with its higher Brier Score (BS).

Table \ref{tab:metrics} presents the classification metrics computed on the holdout dataset for the four selected models. The models induced from the CHEM, FEATENG, and FEATENG+GS datasets yielded similar performance, achieving an F1-Score of 0.88 and accuracy of 0.82. Although not excellent, these metrics demonstrate that these models are useful, especially considering that predicting glass formation from a liquid is a very difficult modeling problem. The model induced from the FEATENG dataset may be preferable due to its lower complexity (20 input features versus 67 for the CHEM model).

The model induced from the GS dataset (using only glass stability parameters plus Jezica as features) performed the worst. Unfortunately, the compound uncertainty of predicting the GS dataset features proved too great to induce a good model. This result supports the observations of Allec and co-authors \cslcitation{10}{[10]}.

Given the GS dataset results, it is unsurprising that the FEATENG+GS dataset did not yield any improvement over the FEATENG dataset in terms of performance. One could argue that the model induced from the FEATENG+GS dataset has the advantage of lower complexity (with significantly fewer estimators, see Table \ref{tab:hp_search_space}). However, a question that will not be resolved in this communication remains: which model performs better when extrapolating to new data, the model with higher complexity or the model with lower complexity but reliant on predicted features?

A key challenge in GFA modeling is how to include structural data as material descriptors (such as Q distribution). Structural information is likely the missing piece for improving model performance. However, if included as features, any new inferences would require this information, which may significantly restrict the possibilities for exploiting the induced model.
\subsection{Explainable AI}
\label{sec:org39efa3a}

\begin{figure}
\centering
\begin{subfigure}[h]{0.4\textwidth}
\includegraphics[keepaspectratio,width=\textwidth]{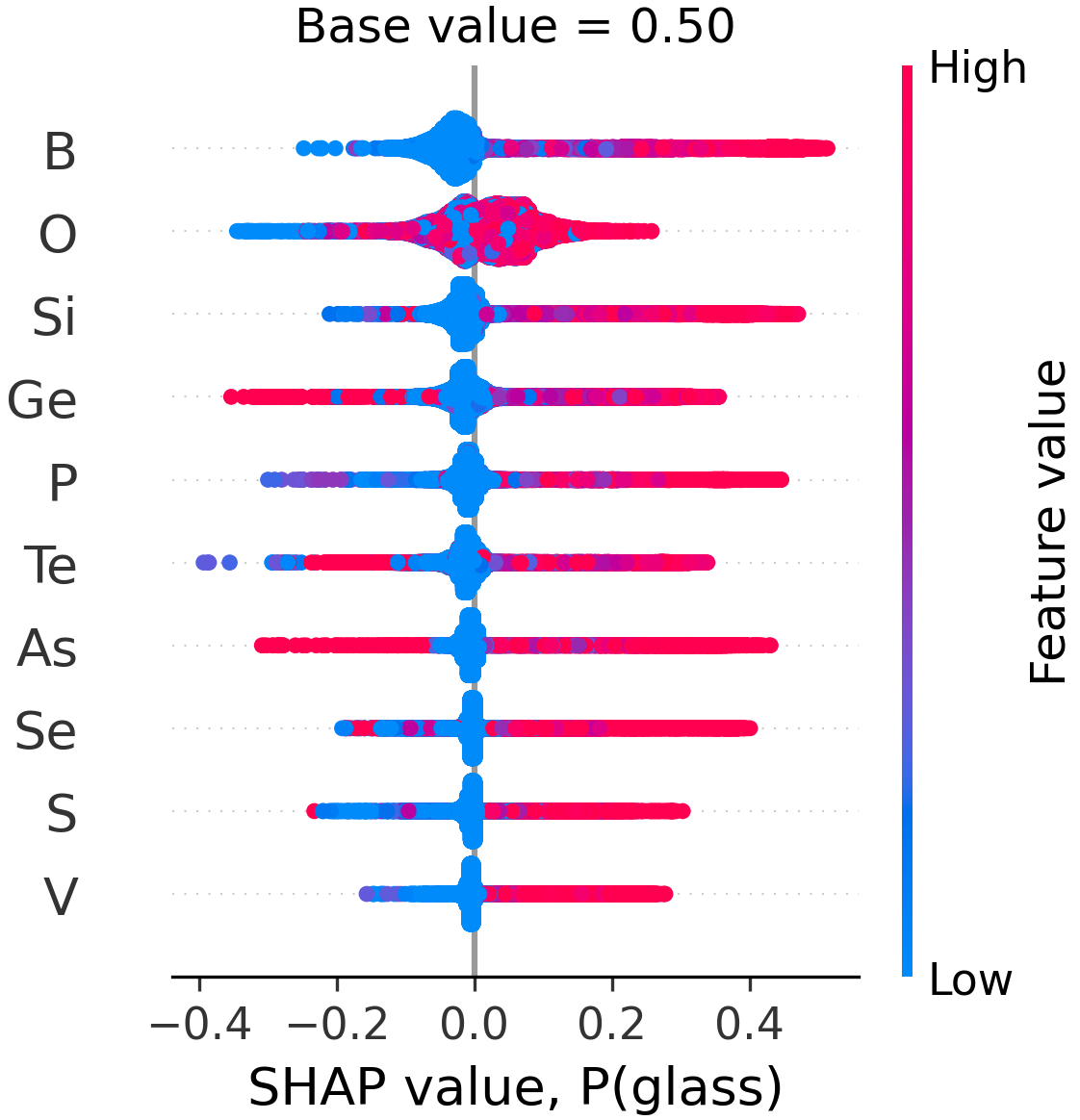}
\caption{}\end{subfigure}

\begin{subfigure}[h]{0.4\textwidth}
\includegraphics[keepaspectratio,width=\textwidth]{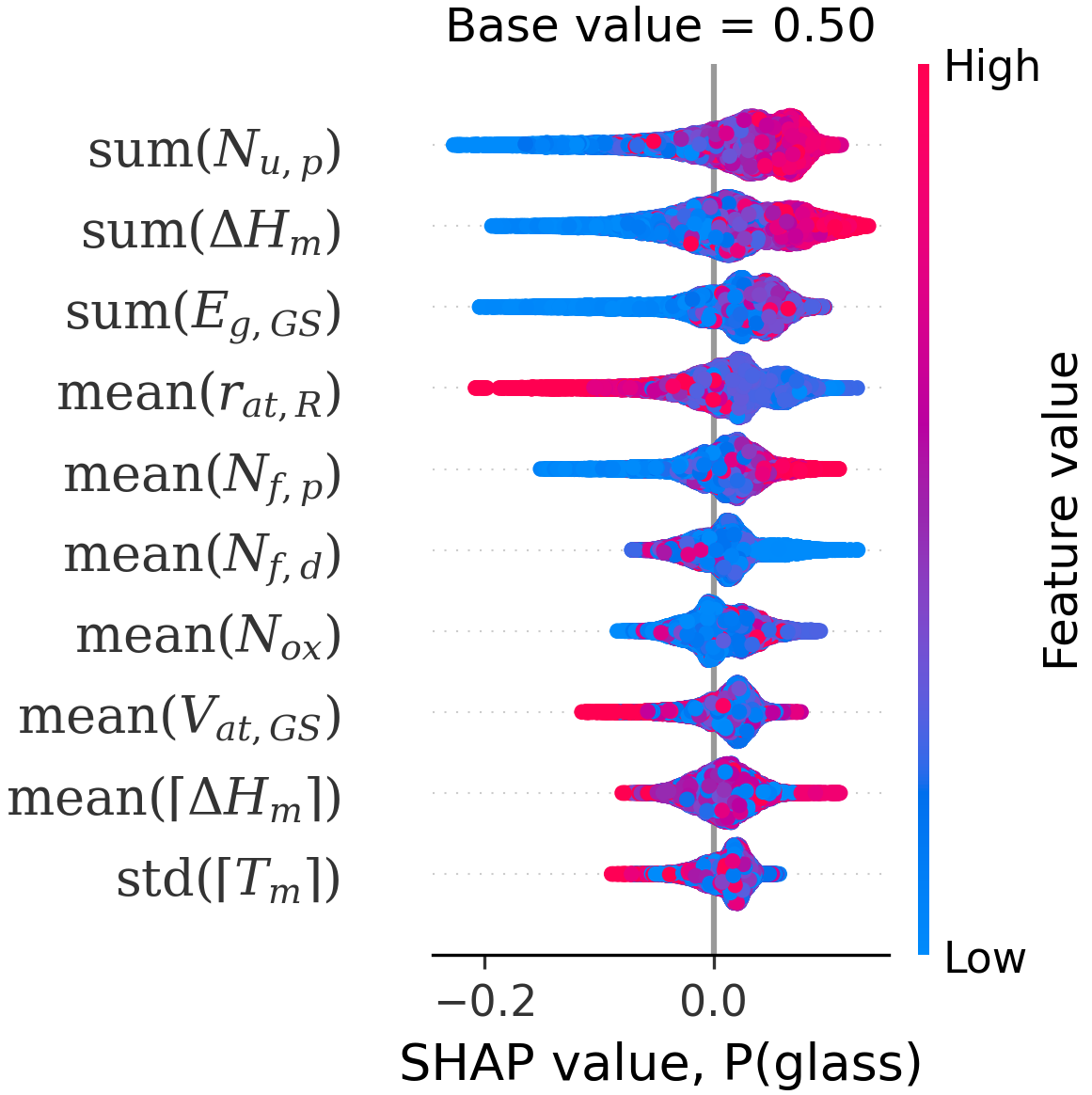}
\caption{}\end{subfigure}

\begin{subfigure}[h]{0.4\textwidth}
\includegraphics[keepaspectratio,width=\textwidth]{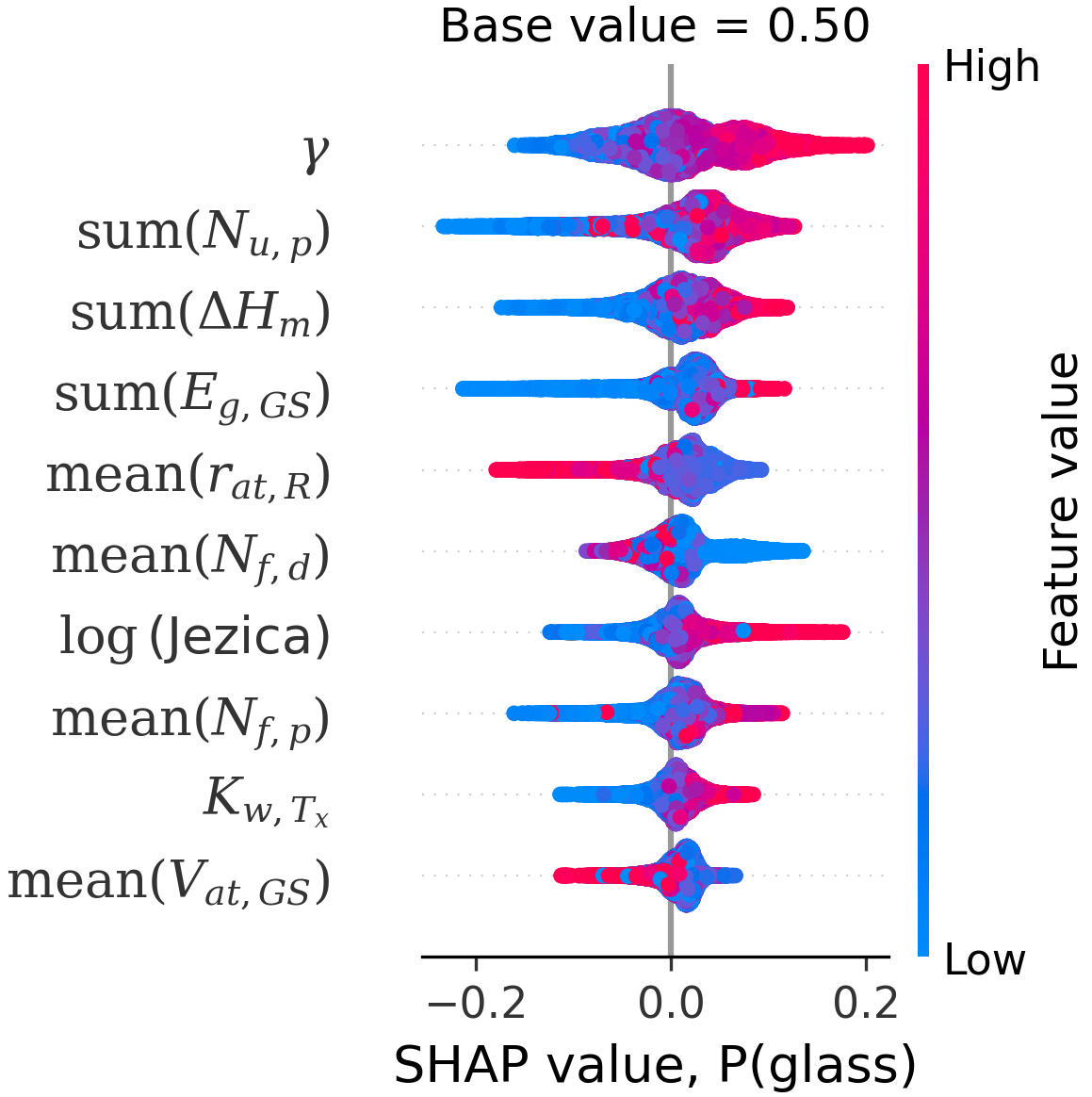}
\caption{}\end{subfigure}
\caption{\label{fig:shap_all}Beeswarm plots of the SHAP values for the 10 most important features of models trained on the (a) CHEM, (b) FEATENG, and (c) FEATENG+GS datasets. Each SHAP value represents the contribution of a feature to the predicted probability of glass formation (\(P(\text{glass})\)). Features marked with the \(\left\lceil \cdot\right\rceil\) operator correspond to absolute descriptors.}
\end{figure}

Fig. \ref{fig:shap_all}a shows the violin plot of SHAP values for the 10 most important features in the CHEM dataset (where importance is defined as the highest mean absolute SHAP value). The plot reveals that increasing the molar fraction of boron, oxygen, silicon, phosphorus, sulfur, and vanadium improves the probability of glass formation. This finding aligns well with established glass science: boron, silicon, phosphorus, and vanadium are well-known glass former cations \cslcitation{46}{[46]}, so their positive contribution is expected. Oxygen forms the backbone of oxide glass networks by connecting glass former elements together. Changes in oxygen molar fraction directly affect the ratio of bridging to non-bridging oxygens, which influences the network's resistance to crystallization.

Sulfur presents an interesting case. In oxide glasses, unlike typical network formers or modifiers, sulfur creates ``non-network oxygen'' \cslcitation{47}{[47]}, individual sulfate groups surrounded by modifier shells that remain outside the glass network. These sulfate groups promote network repolymerization by increasing $Q^3$ and $Q^4$ species \cslcitation{47}{[47]}, which we hypothesize explains why sulfur addition enhances glass formation in oxide glasses. Beyond oxide systems, sulfur also plays a crucial role as a network cation in chalcogenide glasses.

Surprisingly, the glass formers germanium, tellurium, arsenic, and selenium show a no clear pattern in Fig. \ref{fig:shap_all}a. We expected these elements to clearly improve glass formation probability with increasing concentration. This ambiguous behavior may stem from spurious correlations learned by the model (as noted, the model performs well but not perfectly). Alternatively, these elements may exhibit stronger dependence on their chemical environment, producing complex patterns when examined in isolation as in Fig. \ref{fig:shap_all}a.

We found no other clear patterns among the 20 most important chemical elements in the CHEM dataset. 

Fig. \ref{fig:shap_all}b shows the violin plot of SHAP values for the 10 most important features in the FEATENG dataset. Of these features, 8 are weighted descriptors and only 2 are absolute descriptors. This dominance of weighted features reflects their functional role: weighted descriptors perform the actual classification, while absolute descriptors primarily serve for data clustering.

The list below categorizes these features by their correlation with glass formation probability: positive correlation (indicated by ``\(\uparrow\)''), negative correlation (indicated by ``\(\downarrow\)''), and unclear correlation (indicated by ``\textasciitilde{}'').

\begin{itemize}
\item (\(\uparrow\)) weighted sum of the number of p unfilled valence orbitals
\item (\(\uparrow\)) weighted sum of melting enthalpy
\item (\(\uparrow\)) weighted sum of the DFT bandgap energy of the \(T=0\) K ground state
\item (\(\uparrow\)) weighted mean of the number of filled p valence orbitals
\item (\(\downarrow\)) weighted mean of the atomic radius
\item (\(\downarrow\)) absolute std. dev. of the melting point
\item (\textasciitilde{}) weighted mean of the number of filled d valence orbitals
\item (\textasciitilde{}) weighted mean of the number of oxidation states
\item (\textasciitilde{}) weighted mean of the DFT volume of the \(T=0\) K ground state
\item (\textasciitilde{}) absolute mean of melting enthalpy
\end{itemize}

The positive correlation with DFT bandgap energy is particularly noteworthy, suggesting that electronic structure calculations could serve as predictive tools for screening glass-forming compositions. The negative correlation with melting point standard deviation indicates that compositional homogeneity matters; significant differences in the thermal stability of constituent components hinder glass formation.

Interestingly, atomic size mismatch does not appear among the 10 most important features identified by SHAP analysis, despite being included in the 20 selected physicochemical features of the FEATENG dataset. Instead, mean atomic radius proved more impactful for glass formation than size disparity. We attribute this tendency of larger elements to favor glass formation to their slower kinetics, which hinder crystallization.

Features with unclear correlation (those marked with ``\textasciitilde{}'') likely participate in higher-order interactions with other features. For example, the role of d orbitals in glass formation may depend on the specific chemical environment in complex ways not captured by examining this feature in isolation.

Overall, the SHAP analysis identifies electronic structure, melting properties, and atomic radius as the most important features for glass formation. While the physicochemical feature model shows no significant performance improvement over the atomic fraction model, we expect it to be more resilient when extrapolating beyond the training compositional space. The disadvantage is that, in this scenario, modifying composition to improve glass formation is not straightforward, as compositional changes affect most (if not all) physicochemical features simultaneously. However, this limitation can be circumvented by using optimization algorithms (with constraints if needed) to search the compositional space.

Fig. \ref{fig:shap_all}c shows the violin plot of SHAP values for the 10 most important features in the FEATENG+GS dataset. In addition to physicochemical features, this dataset includes glass stability parameters and Jezica. Among the features shown in Fig. \ref{fig:shap_all}c, we observe \(\gamma\), Jezica, and \(K_{w,T_x}\) all exhibiting positive correlation with glass formation.

Notably, the most important feature in this analysis is the Lu and Liu \(\gamma\) parameter, which was ranked as the third-best glass stability parameter for predicting glass-forming ability in a previous study \cslcitation{8}{[8]}. Jezica and \(K_{w,T_c}\) appeared at lower positions in that ranking. Jezica is unique in that it does not require crystallization information, while \(K_{w,T_c}\) was identified as the best parameter for predicting GFA in prior work \cslcitation{8}{[8]}.

Given the significantly worse performance of the model trained on the GS dataset, investigating the correlations it learned in detail is inadvisable, as they likely do not reflect actual causal relationships in nature. 

Finally, it is worth emphasizing that SHAP analysis reveals correlations learned by the model from the available data, and as the well-known maxim states, \emph{correlation does not imply causation}. These results should therefore be interpreted as insights to guide further investigation rather than as established causal relationships.
\subsection{Open science}
\label{sec:orgb94094d}

The codebase for this work is hosted in a Git repository at \url{https://github.com/drcassar/gfa-models}, and the binarized trained models are available in a Zenodo repository \cslcitation{48}{[48]}.

We named the collection of models presented in this work VITRIFY, which stands for \emph{Vitrification Inference Tool for Rapid Identification of glass-Forming abilitY}. VITRIFY is also available in version 0.6.0 of the Python module \texttt{GlassPy} \cslcitation{23}{[23]} for ease of use.
\section{Summary and conclusion}
\label{sec:orgcab6d50}

While this work does not solve the nature of glass and the glass transition, it represents a step toward predicting whether an inorganic nonmetallic liquid will form a glass under typical laboratory conditions. Following George Box's principle that ``all models are wrong, but some are useful,'' we present useful and well-calibrated models that can predict the probability of glass formation, trained in inorganic nonmetallic liquids containing up to 8 different chemical elements.

The primary application of these models is inverse design of new glasses. A fast and reasonably reliable method for predicting glass formation can help navigate the vast compositional space accessible to glasses, which is not constrained by crystalline stoichiometric rules.

Furthermore, explainable AI techniques reinforced existing knowledge in the field and enabled us to obtain empirical insights into glass formation mechanisms. For example, we observed that the average bandgap energy of glass constituents (obtained by DFT) is positively correlated with the probability of glass formation.

Glass stability parameters and Jezica alone were insufficient to induce a good predictive model. When combined with physicochemical descriptors, these parameters do not improve performance but substantially reduce model complexity, indicating they provide relevant information. However, these parameters were calculated from predicted values of \(T_g\), \(T_\mathrm{liq}\), \(T_c\), \(T_x\), and \(\eta(T_\mathrm{liq})\), a process known to yield high uncertainty due to compounding errors. We hypothesize that improving the prediction of these parameters may enhance model performance, though likely not substantially. The current challenge is incorporating structural information in a computationally efficient manner that preserves the model's utility for composition screening.

The obtained models obtained in this work are named VITRIFY and are free software, available to the community in the Python module \texttt{GlassPy}. 
\section*{Acknowledgements}
\label{sec:orgc60e84d}
This study was financed in part by the Coordenação de Aperfeiçoamento de Pessoal de Nível Superior - Brasil (CAPES) - Finance Code 001. DRC and DPLC acknowledge the funding from CNPq - INCT (National Institute of Science and Technology on Materials Informatics, grant n. 371610/2023-0). DRC acknowledges the financial support from the São Paulo Research Foundation (FAPESP) through the Research, Innovation and Dissemination Center for Molecular Engineering for Advanced Materials – CEMol (Grant CEPID No. 2024/00989-7), and through the Thematic Project Grant (No. 2023/09820-2). ACBL and DPLC acknowledge the Ilum School of Science (Brazilian Center for Research in Energy and Materials) for the funding and the structural availability of the high-performance computing clusters.
\section*{References}
\label{sec:org95df3bf}
 \begingroup \leftskip=17pt \parindent=-\leftskip
\begin{cslbibliography}{0}{0}
\cslbibitem{1}{\cslleftmargin{[1]}\cslrightinline{E. D. Zanotto and J. C. Mauro, “The glassy state of matter: Its definition and ultimate fate,” \textit{Journal of non-crystalline solids}, vol. 471, pp. 490–495, 2017, doi: \href{https://doi.org/10.1016/j.jnoncrysol.2017.05.019}{10.1016/j.jnoncrysol.2017.05.019}.}}

\cslbibitem{2}{\cslleftmargin{[2]}\cslrightinline{H. Weintraub \textit{et al.}, “Through the Glass Lightly,” \textit{Science}, vol. 267, no. 5204, pp. 1609–1618, 1995, Accessed: Jan. 20, 2015. [Online]. Available: \url{https://www.jstor.org/stable/2886736}}}

\cslbibitem{3}{\cslleftmargin{[3]}\cslrightinline{E. D. Zanotto and D. R. Cassar, “The microscopic origin of the extreme glass-forming ability of Albite and B2O3,” \textit{Scientific reports}, vol. 7, p. 43022, 2017, doi: \href{https://doi.org/10.1038/srep43022}{10.1038/srep43022}.}}

\cslbibitem{4}{\cslleftmargin{[4]}\cslrightinline{C. Suryanarayana and A. Inoue, \textit{Bulk Metallic Glasses}. CRC Press, 2011.}}

\cslbibitem{5}{\cslleftmargin{[5]}\cslrightinline{J. Jiusti, E. D. Zanotto, D. R. Cassar, and M. R. B. Andreeta, “Viscosity and liquidus-based predictor of glass-forming ability of oxide glasses,” \textit{Journal of the american ceramic society}, vol. 103, no. 2, pp. 921–932, 2020, doi: \href{https://doi.org/10.1111/jace.16732}{10.1111/jace.16732}.}}

\cslbibitem{6}{\cslleftmargin{[6]}\cslrightinline{M. L. F. Nascimento, L. Souza, E. B. Ferreira, and E. D. Zanotto, “Can glass stability parameters infer glass forming ability?,” \textit{Journal of non-crystalline solids}, vol. 351, no. 40-42, pp. 3296–3308, 2005, doi: \href{https://doi.org/10.1016/j.jnoncrysol.2005.08.013}{10.1016/j.jnoncrysol.2005.08.013}.}}

\cslbibitem{7}{\cslleftmargin{[7]}\cslrightinline{S. Liu, H. Tao, Y. Zhang, and Y. Yue, “A new approach for determining the critical cooling rates of nucleation in glass-forming liquids,” \textit{Journal of the american ceramic society}, 2017, doi: \href{https://doi.org/10.1111/jace.14926}{10.1111/jace.14926}.}}

\cslbibitem{8}{\cslleftmargin{[8]}\cslrightinline{J. Jiusti, D. R. Cassar, and E. D. Zanotto, “Which glass stability parameters can assess the glass-forming ability of oxide systems?,” \textit{International journal of applied glass science}, vol. 11, no. 4, pp. 612–621, 2020, doi: \href{https://doi.org/10.1111/ijag.15416}{10.1111/ijag.15416}.}}

\cslbibitem{9}{\cslleftmargin{[9]}\cslrightinline{C. J. Wilkinson, C. Trivelpiece, R. Hust, R. S. Welch, S. A. Feller, and J. C. Mauro, “Hybrid machine learning/physics-based approach for predicting oxide glass-forming ability,” \textit{Acta materialia}, vol. 222, p. 117432, 2022, doi: \href{https://doi.org/10.1016/j.actamat.2021.117432}{10.1016/j.actamat.2021.117432}.}}

\cslbibitem{10}{\cslleftmargin{[10]}\cslrightinline{S. I. Allec \textit{et al.}, “Evaluation of GlassNet for physics-informed machine learning of glass stability and glass-forming ability,” \textit{Journal of the american ceramic society}, vol. 107, no. 12, pp. 7784–7799, 2024, doi: \href{https://doi.org/10.1111/jace.19937}{10.1111/jace.19937}.}}

\cslbibitem{11}{\cslleftmargin{[11]}\cslrightinline{R. B. Rosante, R. F. Lancelotti, and E. D. Zanotto, “JeZCA and JeZiCA – Powerful predictors of glass-forming ability,” \textit{Scripta materialia}, vol. 275, p. 117167, 2026, doi: \href{https://doi.org/10.1016/j.scriptamat.2026.117167}{10.1016/j.scriptamat.2026.117167}.}}

\cslbibitem{12}{\cslleftmargin{[12]}\cslrightinline{S. Mukherjee, J. Schroers, Z. Zhou, W. L. Johnson, and W.-K. Rhim, “Viscosity and specific volume of bulk metallic glass-forming alloys and their correlation with glass forming ability,” \textit{Acta materialia}, vol. 52, no. 12, pp. 3689–3695, 2004.}}

\cslbibitem{13}{\cslleftmargin{[13]}\cslrightinline{O. N. Senkov, “Correlation between fragility and glass-forming ability of metallic alloys,” \textit{Physical review b}, vol. 76, no. 10, p. 104202, 2007.}}

\cslbibitem{14}{\cslleftmargin{[14]}\cslrightinline{S. Sharma, R. Vaidyanathan, and C. Suryanarayana, “Criterion for predicting the glass-forming ability of alloys,” \textit{Applied physics letters}, vol. 90, no. 11, p. 111915, 2007, doi: \href{https://doi.org/10.1063/1.2713867}{10.1063/1.2713867}.}}

\cslbibitem{15}{\cslleftmargin{[15]}\cslrightinline{D. Xu, B. D. Wirth, J. Schroers, and W. L. Johnson, “Calculating glass-forming ability in absence of key kinetic and thermodynamic parameters,” \textit{Applied physics letters}, vol. 97, no. 2, p. 024102, 2010, doi: \href{https://doi.org/10.1063/1.3462315}{10.1063/1.3462315}.}}

\cslbibitem{16}{\cslleftmargin{[16]}\cslrightinline{R. M. Forrest and A. L. Greer, “Evolutionary design of machine-learning-predicted bulk metallic glasses,” \textit{Digital discovery}, 2023, doi: \href{https://doi.org/10.1039/D2DD00078D}{10.1039/D2DD00078D}.}}

\cslbibitem{17}{\cslleftmargin{[17]}\cslrightinline{A. Ghorbani, A. Askari, M. Malekan, and M. Nili-Ahmadabadi, “Thermodynamically-guided machine learning modelling for predicting the glass-forming ability of bulk metallic glasses,” \textit{Scientific reports}, vol. 12, no. 1, p. 11754, 2022, doi: \href{https://doi.org/10.1038/s41598-022-15981-2}{10.1038/s41598-022-15981-2}.}}

\cslbibitem{18}{\cslleftmargin{[18]}\cslrightinline{T. Zhang, Z. Long, L. Peng, and Z. Li, “Prediction of glass forming ability of bulk metallic glasses based on convolutional neural network,” \textit{Journal of non-crystalline solids}, vol. 595, p. 121846, 2022, doi: \href{https://doi.org/10.1016/j.jnoncrysol.2022.121846}{10.1016/j.jnoncrysol.2022.121846}.}}

\cslbibitem{19}{\cslleftmargin{[19]}\cslrightinline{G. Liu \textit{et al.}, “Machine learning versus human learning in predicting glass-forming ability of metallic glasses,” \textit{Acta materialia}, vol. 243, p. 118497, 2023, doi: \href{https://doi.org/10.1016/j.actamat.2022.118497}{10.1016/j.actamat.2022.118497}.}}

\cslbibitem{20}{\cslleftmargin{[20]}\cslrightinline{T. Long, Z. Long, and Z. Peng, “Rational design and glass-forming ability prediction of bulk metallic glasses via interpretable machine learning,” \textit{Journal of materials science}, vol. 58, no. 21, pp. 8833–8844, 2023, doi: \href{https://doi.org/10.1007/s10853-023-08528-x}{10.1007/s10853-023-08528-x}.}}

\cslbibitem{21}{\cslleftmargin{[21]}\cslrightinline{J. Verma, P. Bohane, J. Bhatt, and A. K. Srivastav, “Unveiling glass forming ability patterns in bulk metallic glasses via advanced machine learning approaches,” \textit{Journal of non-crystalline solids}, vol. 624, p. 122710, 2024, doi: \href{https://doi.org/10.1016/j.jnoncrysol.2023.122710}{10.1016/j.jnoncrysol.2023.122710}.}}

\cslbibitem{22}{\cslleftmargin{[22]}\cslrightinline{“Epam/SciGlass,” 2019, \textit{EPAM Systems}.}}

\cslbibitem{23}{\cslleftmargin{[23]}\cslrightinline{D. Cassar, “GlassPy,” 2026, \textit{Zenodo}. doi: \href{https://doi.org/10.5281/zenodo.19006369}{10.5281/zenodo.19006369}.}}

\cslbibitem{24}{\cslleftmargin{[24]}\cslrightinline{D. R. Cassar, “GlassNet: A multitask deep neural network for predicting many glass properties,” \textit{Ceramics international}, vol. 49, no. 22, Part B, pp. 36013–36024, 2023, doi: \href{https://doi.org/10.1016/j.ceramint.2023.08.281}{10.1016/j.ceramint.2023.08.281}.}}

\cslbibitem{25}{\cslleftmargin{[25]}\cslrightinline{F. J. Ferri, P. Pudil, M. Hatef, and J. Kittler, “Comparative study of techniques for large-scale feature selection,” in \textit{Machine intelligence and pattern recognition}, vol. 16, Elsevier, 1994, pp. 403–413. doi: \href{https://doi.org/10.1016/B978-0-444-81892-8.50040-7}{10.1016/B978-0-444-81892-8.50040-7}.}}

\cslbibitem{26}{\cslleftmargin{[26]}\cslrightinline{M. C. Weinberg, “Glass-forming ability and glass stability in simple systems,” \textit{Journal of non-crystalline solids}, vol. 167, no. 1, pp. 81–88, 1994, doi: \href{https://doi.org/10.1016/0022-3093(94)90370-0}{10.1016/0022-3093(94)90370-0}.}}

\cslbibitem{27}{\cslleftmargin{[27]}\cslrightinline{A. Hrub\{\`y\}, “Evaluation of glass-forming tendency by means of DTA,” \textit{Czechoslovak journal of physics b}, vol. 22, no. 11, pp. 1187–1193, 1972.}}

\cslbibitem{28}{\cslleftmargin{[28]}\cslrightinline{M. Saad and M. Poulain, “Glass forming ability criterion,” in \textit{Materials science forum}, Trans Tech Publ, 1987, pp. 11–18.}}

\cslbibitem{29}{\cslleftmargin{[29]}\cslrightinline{Z. P. Lu and C. T. Liu, “A new glass-forming ability criterion for bulk metallic glasses,” \textit{Acta materialia}, vol. 50, no. 13, pp. 3501–3512, 2002.}}

\cslbibitem{30}{\cslleftmargin{[30]}\cslrightinline{J. S. Bergstra, R. Bardenet, Y. Bengio, and B. Kégl, “Algorithms for hyper-parameter optimization,” in \textit{Advances in Neural Information Processing Systems}, 2011, pp. 2546–2554.}}

\cslbibitem{31}{\cslleftmargin{[31]}\cslrightinline{T. Akiba, S. Sano, T. Yanase, T. Ohta, and M. Koyama, “Optuna: A next-generation hyperparameter optimization framework,” in \textit{Proceedings of the 25th ACM SIGKDD international conference on knowledge discovery and data mining}, 2019.}}

\cslbibitem{32}{\cslleftmargin{[32]}\cslrightinline{L. S. Shapley, “A value for n-person games,” 1953.}}

\cslbibitem{33}{\cslleftmargin{[33]}\cslrightinline{S. M. Lundberg \textit{et al.}, “From local explanations to global understanding with explainable AI for trees,” \textit{Nature machine intelligence}, vol. 2, no. 1, pp. 56–67, 2020, doi: \href{https://doi.org/10.1038/s42256-019-0138-9}{10.1038/s42256-019-0138-9}.}}

\cslbibitem{34}{\cslleftmargin{[34]}\cslrightinline{S. M. Lundberg \textit{et al.}, “Explainable machine-learning predictions for the prevention of hypoxaemia during surgery,” \textit{Nature biomedical engineering}, vol. 2, no. 10, pp. 749–760, 2018, doi: \href{https://doi.org/10.1038/s41551-018-0304-0}{10.1038/s41551-018-0304-0}.}}

\cslbibitem{35}{\cslleftmargin{[35]}\cslrightinline{S. M. Lundberg and S.-I. Lee, “A unified approach to interpreting model predictions,” \textit{Advances in neural information processing systems}, vol. 30, pp. 4765–4774, 2017.}}

\cslbibitem{36}{\cslleftmargin{[36]}\cslrightinline{A. L. Greer, “Confusion by design,” \textit{Nature}, vol. 366, no. 6453, pp. 303–304, 1993, doi: \href{https://doi.org/10.1038/366303a0}{10.1038/366303a0}.}}

\cslbibitem{37}{\cslleftmargin{[37]}\cslrightinline{V. M. Goldschmidt, “Die Gesetze der Krystallochemie,” \textit{Die Naturwissenschaften}, vol. 14, no. 21, pp. 477–485, 1926, doi: \href{https://doi.org/10.1007/BF01507527}{10.1007/BF01507527}.}}

\cslbibitem{38}{\cslleftmargin{[38]}\cslrightinline{W. H. Zachariasen, “The Atomic Arrangement in Glass,” \textit{Journal of the american chemical society}, vol. 54, no. 10, pp. 3841–3851, 1932, doi: \href{https://doi.org/10.1021/ja01349a006}{10.1021/ja01349a006}.}}

\cslbibitem{39}{\cslleftmargin{[39]}\cslrightinline{K.-H. Sun, “Fundamental Condition of Glass Formation,” \textit{Journal of the american ceramic society}, vol. 30, no. 9, pp. 277–281, 1947, doi: \href{https://doi.org/10.1111/j.1151-2916.1947.tb19654.x}{10.1111/j.1151-2916.1947.tb19654.x}.}}

\cslbibitem{40}{\cslleftmargin{[40]}\cslrightinline{A. Dietzel, “Die Kationenfeldstärken und ihre Beziehungen zu Entglasungsvorgängen, zur Verbindungsbildung und zu den Schmelzpunkten von Silicaten,” \textit{Zeitschrift für elektrochemie und angewandte physikalische chemie}, vol. 48, no. 1, pp. 9–23, 1942, doi: \href{https://doi.org/10.1002/bbpc.19420480104}{10.1002/bbpc.19420480104}.}}

\cslbibitem{41}{\cslleftmargin{[41]}\cslrightinline{J. C. Mauro, “Topological constraint theory of glass,” \textit{American ceramic society bulletin}, vol. 90, no. 4, p. 31, 2011.}}

\cslbibitem{42}{\cslleftmargin{[42]}\cslrightinline{J. C. Phillips and M. F. Thorpe, “Constraint theory, vector percolation and glass formation,” \textit{Solid state communications}, vol. 53, no. 8, pp. 699–702, 1985, doi: \href{https://doi.org/10.1016/0038-1098(85)90381-3}{10.1016/0038-1098(85)90381-3}.}}

\cslbibitem{43}{\cslleftmargin{[43]}\cslrightinline{J. C. Phillips, “Topology of covalent non-crystalline solids I: Short-range order in chalcogenide alloys,” \textit{Journal of non-crystalline solids}, vol. 34, no. 2, pp. 153–181, 1979, doi: \href{https://doi.org/10.1016/0022-3093(79)90033-4}{10.1016/0022-3093(79)90033-4}.}}

\cslbibitem{44}{\cslleftmargin{[44]}\cslrightinline{D. R. Cassar, “Crystallization driving force of supercooled oxide liquids,” \textit{International journal of applied glass science}, vol. 7, no. 3, pp. 262–269, 2016, doi: \href{https://doi.org/10.1111/ijag.12218}{10.1111/ijag.12218}.}}

\cslbibitem{45}{\cslleftmargin{[45]}\cslrightinline{J. W. Schmelzer and A. S. Abyzov, “Crystallization of glass-forming liquids: Thermodynamic driving force,” \textit{Journal of non-crystalline solids}, vol. 449, pp. 41–49, 2016, doi: \href{https://doi.org/10.1016/j.jnoncrysol.2016.07.005}{10.1016/j.jnoncrysol.2016.07.005}.}}

\cslbibitem{46}{\cslleftmargin{[46]}\cslrightinline{A. K. Varshneya and J. C. Mauro, \textit{Fundamentals of Inorganic Glasses}, 3 edition. Elsevier, 2019.}}

\cslbibitem{47}{\cslleftmargin{[47]}\cslrightinline{S. Vaishnav, A. C. Hannon, E. R. Barney, and P. A. Bingham, “Neutron Diffraction and Raman Studies of the Incorporation of Sulfate in Silicate Glasses,” \textit{The journal of physical chemistry c}, vol. 124, no. 9, pp. 5409–5424, 2020, doi: \href{https://doi.org/10.1021/acs.jpcc.9b10924}{10.1021/acs.jpcc.9b10924}.}}

\cslbibitem{48}{\cslleftmargin{[48]}\cslrightinline{D. P. d. L. Carvalho and D. Cassar, “Glass Formation Classifier: ML Models for Inorganic Nonmetallic Systems,” 2026, doi: \href{https://doi.org/10.5281/zenodo.18964978}{10.5281/zenodo.18964978}.}}

\cslbibitem{49}{\cslleftmargin{[49]}\cslrightinline{“Scikit-learn documentation of RandomForestClassifier,” \textit{https://scikit-learn.org/1.8/modules/generated/sklearn.ensemble.RandomForestClassifier.html}.}}

\cslbibitem{50}{\cslleftmargin{[50]}\cslrightinline{M. Rahm, R. Hoffmann, and N. W. Ashcroft, “Corrigendum: Atomic and Ionic Radii of Elements 1–96,” \textit{Chemistry – a european journal}, vol. 23, no. 16, p. 4017, 2017, doi: \href{https://doi.org/10.1002/chem.201700610}{10.1002/chem.201700610}.}}

\cslbibitem{51}{\cslleftmargin{[51]}\cslrightinline{M. Rahm, R. Hoffmann, and N. W. Ashcroft, “Atomic and Ionic Radii of Elements 1–96,” \textit{Chemistry – a european journal}, vol. 22, no. 41, pp. 14625–14632, 2016, doi: \href{https://doi.org/10.1002/chem.201602949}{10.1002/chem.201602949}.}}

\cslbibitem{52}{\cslleftmargin{[52]}\cslrightinline{B. Cordero \textit{et al.}, “Covalent radii revisited,” \textit{Dalton transactions}, no. 21, pp. 2832–2838, 2008, doi: \href{https://doi.org/10.1039/B801115J}{10.1039/B801115J}.}}

\cslbibitem{53}{\cslleftmargin{[53]}\cslrightinline{A. K. Rappe, C. J. Casewit, K. S. Colwell, W. A. Goddard, and W. M. Skiff, “UFF, a full periodic table force field for molecular mechanics and molecular dynamics simulations,” \textit{Journal of the american chemical society}, vol. 114, no. 25, pp. 10024–10035, 1992, doi: \href{https://doi.org/10.1021/ja00051a040}{10.1021/ja00051a040}.}}

\cslbibitem{54}{\cslleftmargin{[54]}\cslrightinline{H. Glawe, A. Sanna, E. K. U. Gross, and M. A. L. Marques, “The optimal one dimensional periodic table: A modified Pettifor chemical scale from data mining,” \textit{New journal of physics}, vol. 18, no. 9, p. 093011, 2016, doi: \href{https://doi.org/10.1088/1367-2630/18/9/093011}{10.1088/1367-2630/18/9/093011}.}}

\cslbibitem{55}{\cslleftmargin{[55]}\cslrightinline{D. G. Pettifor, “A chemical scale for crystal-structure maps,” \textit{Solid state communications}, vol. 51, no. 1, pp. 31–34, 1984, doi: \href{https://doi.org/10.1016/0038-1098(84)90765-8}{10.1016/0038-1098(84)90765-8}.}}

\cslbibitem{56}{\cslleftmargin{[56]}\cslrightinline{W. M. Haynes, \textit{CRC Handbook of Chemistry and Physics}. CRC Press, 2014.}}

\cslbibitem{57}{\cslleftmargin{[57]}\cslrightinline{L. Ward, A. Agrawal, A. Choudhary, and C. Wolverton, “A general-purpose machine learning framework for predicting properties of inorganic materials,” \textit{Npj computational materials}, vol. 2, no. 1, pp. 1–7, 2016, doi: \href{https://doi.org/10.1038/npjcompumats.2016.28}{10.1038/npjcompumats.2016.28}.}}

\cslbibitem{58}{\cslleftmargin{[58]}\cslrightinline{Ł. Mentel, “mendeleev – A Python resource for properties of chemical elements, ions and isotopes,” 2014.}}

\cslbibitem{59}{\cslleftmargin{[59]}\cslrightinline{S. S. Batsanov, “Dielectric Methods of Studying the Chemical Bond and the Concept of Electronegativity,” \textit{Russian chemical reviews}, vol. 51, no. 7, p. 684, 1982, doi: \href{https://doi.org/10.1070/RC1982v051n07ABEH002900}{10.1070/RC1982v051n07ABEH002900}.}}

\end{cslbibliography}
 \newpage
 \endgroup

\newpage
\section*{Tables and figures}
\label{sec:orgcb3455e}
\begin{landscape}

\begin{table}[htbp]
\caption{\label{tab:hp_search_space}Hyperparameter search space for the Random Forest Classifier model and the selected values for the four studied datasets. The notation follows scikit-learn version 1.8 \cslcitation{49}{[49]}.}
\centering
\begin{tabular}{llrrrr}
Hyperparameter & Search space & CHEM & FEATENG & GS & FEATENG+GS\\
\hline
n\_estimators & \{10, 11, \ldots{}, 1000\} & 722 & 719 & 419 & 384\\
criterion & \{gini, entropy, log\_loss\} & gini & log\_loss & log\_loss & log\_loss\\
min\_samples\_split & \{2, 3, \ldots{}, 20\} & 3 & 7 & 15 & 8\\
min\_samples\_leaf & \{1, 2, \ldots{}, 20\} & 1 & 2 & 6 & 1\\
max\_depth & \{None\} \(\cup\) \{1, 2, \ldots{}, 100\} & 90 & None & 45 & None\\
max\_features & \{sqrt, log2, None\} \(\cup\) {[}\num{e-5}, 1] & None & sqrt & \num{2.18e-5} & \num{4.97e-1}\\
max\_leaf\_nodes & \{None\} \(\cup\) \{10, 11, \ldots{}, 1000\} & None & None & None & None\\
min\_impurity\_decrease & \{0\} \(\cup\) {[}\num{e-5}, 1] & 0 & 0 & 0 & \num{9.91e-5}\\
class\_weight & \{None, balanced, balanced\_subsample\} & balanced\_subsample & balanced & balanced\_subsample & balanced\\
ccp\_alpha & \{0\} \(\cup\) {[}\num{e-5}, 1] & \num{4.43e-5} & 0 & \num{1.82e-5} & \num{1.30e-4}\\
max\_samples & \{None\} \(\cup\) {[}\num{e-4}, 1] & None & None & None & None\\
\hline
\end{tabular}
\end{table}
\end{landscape}

\newpage

\begin{longtable}{lrrrr}
\caption{\label{tab:descriptive}Descriptive statistics of chemical elements in the entire dataset, before the holdout split.}
\\
Feature & Min & Max & Mean & Std\\
\hline
\endfirsthead
\multicolumn{5}{l}{Continued from previous page} \\
\hline

Feature & Min & Max & Mean & Std \\

\hline
\endhead
\hline\multicolumn{5}{r}{Continued on next page} \\
\endfoot
\endlastfoot
\hline
Ag & 0.000 & 0.666 & 0.011 & 0.054\\
Al & 0.000 & 0.400 & 0.009 & 0.033\\
As & 0.000 & 0.883 & 0.022 & 0.085\\
B & 0.000 & 0.695 & 0.050 & 0.101\\
Ba & 0.000 & 0.391 & 0.010 & 0.034\\
Be & 0.000 & 0.468 & 0.002 & 0.021\\
Bi & 0.000 & 0.400 & 0.010 & 0.043\\
Br & 0.000 & 0.744 & 0.007 & 0.060\\
Ca & 0.000 & 0.395 & 0.006 & 0.028\\
Cd & 0.000 & 0.499 & 0.005 & 0.029\\
Ce & 0.000 & 0.100 & 0.000 & 0.003\\
Cl & 0.000 & 0.750 & 0.006 & 0.053\\
Co & 0.000 & 0.327 & 0.001 & 0.009\\
Cr & 0.000 & 0.183 & 0.000 & 0.005\\
Cs & 0.000 & 0.469 & 0.002 & 0.022\\
Cu & 0.000 & 0.500 & 0.004 & 0.031\\
Dy & 0.000 & 0.146 & 0.000 & 0.003\\
Er & 0.000 & 0.153 & 0.000 & 0.004\\
Eu & 0.000 & 0.344 & 0.000 & 0.004\\
F & 0.000 & 0.792 & 0.031 & 0.133\\
Fe & 0.000 & 0.308 & 0.003 & 0.019\\
Ga & 0.000 & 0.382 & 0.009 & 0.038\\
Gd & 0.000 & 0.249 & 0.001 & 0.008\\
Ge & 0.000 & 1.000 & 0.042 & 0.092\\
H & 0.000 & 0.574 & 0.001 & 0.008\\
Hf & 0.000 & 0.184 & 0.000 & 0.007\\
Hg & 0.000 & 0.449 & 0.001 & 0.013\\
Ho & 0.000 & 0.154 & 0.000 & 0.003\\
I & 0.000 & 0.707 & 0.012 & 0.060\\
In & 0.000 & 0.300 & 0.002 & 0.015\\
K & 0.000 & 0.593 & 0.011 & 0.045\\
La & 0.000 & 0.360 & 0.003 & 0.018\\
Li & 0.000 & 0.556 & 0.011 & 0.046\\
Lu & 0.000 & 0.130 & 0.000 & 0.002\\
Mg & 0.000 & 0.446 & 0.004 & 0.022\\
Mn & 0.000 & 0.362 & 0.002 & 0.017\\
Mo & 0.000 & 0.236 & 0.002 & 0.016\\
N & 0.000 & 0.511 & 0.000 & 0.010\\
Na & 0.000 & 0.666 & 0.015 & 0.054\\
Nb & 0.000 & 0.277 & 0.003 & 0.021\\
Nd & 0.000 & 0.230 & 0.001 & 0.007\\
Ni & 0.000 & 0.308 & 0.000 & 0.006\\
O & 0.000 & 0.750 & 0.433 & 0.270\\
P & 0.000 & 0.903 & 0.023 & 0.061\\
Pb & 0.000 & 0.464 & 0.015 & 0.052\\
Pr & 0.000 & 0.138 & 0.000 & 0.002\\
Rb & 0.000 & 0.469 & 0.001 & 0.017\\
S & 0.000 & 0.985 & 0.047 & 0.154\\
Sb & 0.000 & 0.895 & 0.009 & 0.046\\
Sc & 0.000 & 0.172 & 0.000 & 0.006\\
Se & 0.000 & 1.000 & 0.058 & 0.176\\
Si & 0.000 & 0.649 & 0.025 & 0.064\\
Sm & 0.000 & 0.205 & 0.000 & 0.004\\
Sn & 0.000 & 0.601 & 0.003 & 0.021\\
Sr & 0.000 & 0.396 & 0.004 & 0.024\\
Ta & 0.000 & 0.242 & 0.001 & 0.009\\
Tb & 0.000 & 0.227 & 0.000 & 0.003\\
Te & 0.000 & 1.000 & 0.049 & 0.130\\
Ti & 0.000 & 0.305 & 0.004 & 0.022\\
Tl & 0.000 & 0.664 & 0.006 & 0.039\\
Tm & 0.000 & 0.119 & 0.000 & 0.002\\
V & 0.000 & 0.286 & 0.010 & 0.042\\
W & 0.000 & 0.250 & 0.002 & 0.015\\
Y & 0.000 & 0.398 & 0.001 & 0.010\\
Yb & 0.000 & 0.145 & 0.000 & 0.003\\
Zn & 0.000 & 0.434 & 0.009 & 0.036\\
Zr & 0.000 & 0.272 & 0.002 & 0.014\\
\hline
\end{longtable}

\newpage

\begin{table}[htbp]
\caption{\label{tab:symbols}Symbols and meanings of the physicochemical properties used in this work.}
\centering
\begin{tabular}{ll}
\hline
Symbol & Meaning\\
\hline
\(r_{at,R}\) & Atomic radius in the Rahm scale (pm) \cslcitation{50}{[50]}, \cslcitation{51}{[51]}\\
\(r_{cov,C}\) & Covalent radius in the Cordero scale (pm) \cslcitation{52}{[52]}\\
\(r_{W,UFF}\) & Van der Waals radius in the UFF force field (pm) \cslcitation{53}{[53]}\\
\(n_{Pe}\) & Pettifor number \cslcitation{54}{[54]}, \cslcitation{55}{[55]}\\
\(T_m\) & Melting point (K) \cslcitation{56}{[56]}\\
\(\Delta H_m\) & Melting enthalpy (kJ/mol) \cslcitation{57}{[57]}\\
\(E_{g,GS}\) & DFT bandgap energy of \(T=0\,\mathrm{K}\) ground state (eV) \cslcitation{57}{[57]}\\
\(V_{at,GS}\) & DFT volume per atom of \(T=0\,\mathrm{K}\) ground state \cslcitation{57}{[57]}\\
\(N_{ox}\) & Number of oxidation states \cslcitation{58}{[58]}\\
\(N_{f,s}\) & Number of filled s valence orbitals \cslcitation{57}{[57]}\\
\(N_{f,p}\) & Number of filled p valence orbitals \cslcitation{57}{[57]}\\
\(N_{f,d}\) & Number of filled d valence orbitals \cslcitation{57}{[57]}\\
\(N_{u}\) & Number of unfilled valence orbitals \cslcitation{57}{[57]}\\
\(N_{u,p}\) & Number of unfilled p valence orbitals \cslcitation{57}{[57]}\\
\(N_{u,d}\) & Number of unfilled d valence orbitals \cslcitation{57}{[57]}\\
\(\chi_{MB}\) & Electronegativity in the Martynov--Batsanov scale \cslcitation{59}{[59]}\\
\hline
\end{tabular}
\end{table}

\newpage

\begin{table}[htbp]
\caption{\label{tab:metrics}Classification metrics for the models trained on the 4 datasets studied in this work. Values computed on the holdout dataset. In metrics with an up arrow (\(\uparrow\)), higher values are better. In metrics with a down arrow (\(\downarrow\)), lower values are better. Bold values indicate the best metric for each row.}
\centering
\begin{tabular}{lrrrr}
Dataset & CHEM & FEATENG & GS & FEATENG+GS\\
\hline
Accuracy (\(\uparrow\)) & \textbf{0.826} & 0.824 & 0.704 & 0.822\\
Weighted Accuracy (\(\uparrow\)) & \textbf{0.779} & 0.766 & 0.621 & 0.761\\
ROC-AUC (\(\uparrow\)) & \textbf{0.889} & \textbf{0.889} & 0.695 & 0.886\\
PR-AUC (\(\uparrow\)) & 0.949 & \textbf{0.950} & 0.837 & 0.949\\
Macro PR-AUC (\(\uparrow\)) & 0.865 & \textbf{0.866} & 0.673 & 0.864\\
F1-Score (\(\uparrow\)) & \textbf{0.879} & \textbf{0.879} & 0.797 & 0.878\\
Macro F1-Score (\(\uparrow\)) & \textbf{0.785} & 0.777 & 0.625 & 0.774\\
Brier Score (\(\downarrow\)) & \textbf{0.119} & 0.120 & 0.196 & 0.123\\
Log Loss (\(\downarrow\)) & \textbf{0.373} & 0.377 & 0.577 & 0.385\\
\hline
\end{tabular}
\end{table}

\newpage

\newpage

\newpage

\newpage

\end{document}